\pdftrailerid{} 
\documentclass[conference]{IEEEtran}
\usepackage{graphicx}
\usepackage{subfig}

\usepackage[T1]{fontenc}
\usepackage[scaled=1.0]{DejaVuSansMono}
\usepackage{listings}

\usepackage{textcomp}
\usepackage{multirow}
\usepackage{array}
\usepackage{cite}
\usepackage{url}
\usepackage{balance}
\usepackage{amsmath}
\usepackage{algorithm2e}

\newcolumntype{L}[1]{>{\raggedright\let\newline\\\arraybackslash\hspace{0pt}}m{#1}}
\newcolumntype{C}[1]{>{\centering\let\newline\\\arraybackslash\hspace{0pt}}m{#1}}
\newcolumntype{R}[1]{>{\raggedleft\let\newline\\\arraybackslash\hspace{0pt}}m{#1}}

\lstnewenvironment{code}[1][]%
  {
    \noindent
    \minipage{\linewidth}
    \medskip
    \lstset{basicstyle=\ttfamily\footnotesize,frame=single,#1}}
  {\endminipage}

\makeatletter
\def\footnoterule{\relax%
  \kern-5pt
  \hbox to \columnwidth{\hfill\vrule width 0.5\columnwidth height 0.4pt\hfill}
  \kern4.6pt}
\makeatother

\graphicspath{{images/}}
\DeclareGraphicsExtensions{.pdf,.jpeg,.png,.jpg,.eps}

\renewcommand{\IEEEkeywordsname}{Keywords}

\begin{document}
\title{Performance Measurements of Supercomputing and Cloud Storage Solutions}

\author{\IEEEauthorblockN{Michael Jones,
Jeremy Kepner,
William Arcand,
David Bestor,
Bill Bergeron,
Vijay Gadepally, 
Michael Houle, \\
Matthew Hubbell,
Peter Michaleas,
Andrew Prout,
Albert Reuther,
Siddharth Samsi,
Paul Monticiollo \\
\IEEEauthorblockA{MIT Lincoln Laboratory, Lexington, MA, U.S.A.}}}

\maketitle

\IEEEtitleabstractindextext{%
\begin{abstract}
 Increasing amounts of data from varied sources, particularly in the fields of machine learning and graph analytics, are causing storage requirements to grow rapidly.  A variety of technologies exist for storing and sharing these data, ranging from parallel file systems used by supercomputers to distributed block storage systems found in clouds.  Relatively few comparative measurements exist to inform decisions about which storage systems are best suited for particular tasks.  This work provides these measurements for two of the most popular storage technologies: Lustre and Amazon S3. Lustre is an open-source, high performance, parallel file system used by many of the largest supercomputers in the world.  Amazon's Simple Storage Service, or S3, is part of the Amazon Web Services offering, and offers a scalable, distributed option to store and retrieve data from anywhere on the Internet.   Parallel processing is essential for achieving high performance on modern storage systems.  The performance tests used span the gamut of parallel I/O scenarios, ranging from single-client, single-node Amazon S3 and Lustre performance to a large-scale, multi-client test designed to demonstrate the capabilities of a modern storage appliance under heavy load.   These results show that, when parallel I/O is used correctly (i.e., many simultaneous read or write processes), full network bandwidth performance is achievable and ranged from 10 gigabits/s over a 10 GigE S3 connection to 0.35 terabits/s using Lustre on a 1200 port 10 GigE switch.  These results demonstrate that S3 is well-suited to sharing vast quantities of data over the Internet, while Lustre is well-suited to processing large quantities of data locally.
\end{abstract}

\begin{IEEEkeywords}
High Performance Computing, High Performance Storage, Lustre, Amazon Simple Storage Service, MIT SuperCloud
\end{IEEEkeywords}	
}

\IEEEpeerreviewmaketitle
\IEEEdisplaynontitleabstractindextext

\section{Introduction}
\label{sec:introduction}

\let\thefootnote\relax\footnotetext{This material is based upon work supported by the Defense Advanced Research Projects Agency (DARPA) under Air Force Contract No. FA8721-05-C-0002 and/or FA8702-15-D-0001.  Any opinions, findings, conclusions or recommendations expressed in this material are those of the author(s) and do not necessarily reflect the views of DARPA.}

Rapidly growing diverse data is being collected via varying sensors, social media, and scientific instruments.  These data are being analyzed with machine learning and graph analytics to reveal the complex relationships between different data feeds.  Many machine learning and graph analytics workloads are executed in large data centers on large cached or static data sets.  While research on static and streaming graph analytics problems is primarily limited by the amount of processing required, there exists a growing requirement for high performance and readily available storage systems to manipulate this data in real-time and store it at rest.

A variety of technologies exist for storing and sharing data, ranging from parallel file systems used by supercomputers to distributed block storage systems found in clouds.  Relatively few comparative measurements exist to inform decisions about which storage systems are best suited for particular tasks~\cite{juve2009scientific,juve2010data,yildirim2016parallel}.  This work provides these detailed measurements for two of the most popular storage technologies: Lustre and Amazon S3.

The DARPA Graph Challenge~\cite{darpahive} is an example of a recent effort to push the boundaries of current scalability constraints in the field of graph analytics.  Two challenges have been proposed: a Subgraph Isomorphism challenge~\cite{samsistatic}, which seeks to identify the structure of relationships in a given graph, and a Stochastic Block Partitioning challenge~\cite{kaostreaming} which looks to discover the distinct community structure or specific community membership for each node in the graph.

Large datasets are the "fuel" for this Graph Challenge "rocket," and, today, a variety of curated, public, real-world graphs, as well as synthetic data sets, are available~\cite{challengedatasets}.  The vast majority of these data sets are relatively modest in size by supercomputing standards, measuring tens of gigabytes, but as we move toward achieving the ability to manipulate graphs as large as a trillion edges, the data storage requirements balloon, both in terms of the quantity of storage required and the performance it must deliver in order to manipulate the graphs in near real-time.

This paper outlines the results of performance measurements done as part of the DARPA HIVE~\cite{darpahive} program in which we sought to measure the achievable single-client performance of two classes of storage solutions.  The first class,  supercomputing storage, is represented by the Lustre parallel file system \cite{braam2002lustre} in two back-end network configurations:  Infiniband and 10 Gigabit Ethernet.  The second class,  cloud storage, is represented by the Amazon Simple Storage Service (S3), accessed via a minimal hop-count, low-latency Internet route through a 100-Gigabit backhaul fiber connection on a 10 Gigabit Ethernet-connected client node.  Additionally, we ran similar measurements against a modern high performance compute cluster's storage array in a distributed manner with multiple client nodes to measure obtainable performance in a massively parallel configuration.

The primary goal of this paper is to provide a set of baseline measurements showing the kind of performance actually obtainable from storage systems under real-world conditions.  The organization of the rest of this paper is as follows.  Section \ref{sec:storage} describes the two storage solutions measured in this paper.  Section \ref{sec:environment} provides details on the setup and organization of the experiment as well as detail on our testing methodology.  Section \ref{sec:performance} presents the performance results we obtained during our experiments and an overview of our findings.  Section \ref{sec:summary} summarizes the work and describes future directions.

\section{Data Storage}
\label{sec:storage}

The concept of an object-based storage system was first introduced in the early 1990s by researchers, and within a decade its use was widespread both in industry and high performance computing circles~\cite{factor2005object}.  A storage \textit{object}, much like a \textit{file} in a general-purpose file system, represents a collection of related bytes physically residing on some storage device.  Objects provide a similar abstraction for data access to files and typically offer a standard interface by which permissions, attributes, and other associated metadata can be manipulated~\cite{mesnier2003object}.

\subsection{The Lustre parallel file system}
\label{sec:lustre}

\begin{figure}[ht]
	\centering
	\includegraphics[width=8.75cm,keepaspectratio]{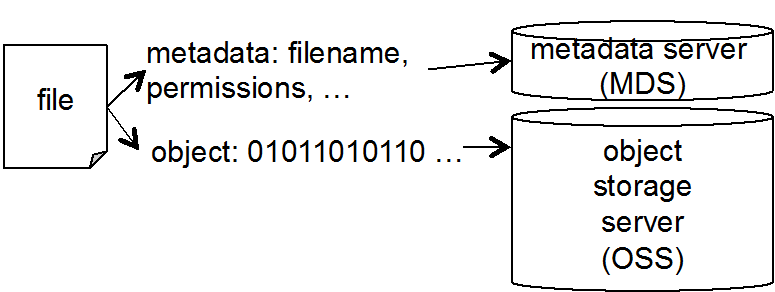}
	\caption{Simplified depiction of the Lustre file system's architecture, split into Metadata Servers (MDS) and Object Storage Servers (OSS).}
	\label{fig:lustrearch}
\end{figure}

Lustre is designed to meet the highest bandwidth file requirements on the largest systems in the world.  The open-source Lustre parallel file system presents itself as a standard POSIX general-purpose file system, and it is mounted by client computers running the Lustre client software.  A file stored in Lustre is broken into two components: metadata and object data (see Figure~\ref{fig:lustrearch}).  Metadata consist of the fields associated with each file such as filename, file permissions, and timestamps.  Object data consist of the binary data stored in the file.  File metadata are stored in the Lustre metadata server (MDS).  Object data are stored in object storage servers (OSS).  (Figure~\ref{fig:lustrearch})  When a client requests data from a file, the MDS returns pointers to the appropriate objects in the OSSes indicating the location of the requested data and the byte ranges in the objects associated with the requested file.  This action is transparent to the user and handled by the Lustre client.  To an application, Lustre operations appear as standard file system operations and require no modification of application code~\cite{braam2003lustre,kepner2015lustre}.

Like most general-purpose file systems, Lustre is designed to maximize the performance of sequential read accesses and not random lookups of data.  This optimization model fits nicely with our proposed use case, as most of the formats typically used to represent large graphs (e.g. Matrix Market I/O \cite{boisvert1996matrix,matrixmarket} and tab or comma-separated values) are mostly composed of large and monolithic streams of ASCII text data.

Lustre's security model relies on standard UNIX permissions, which are simple but flexible enough to accommodate most needs.

\subsection{Amazon Simple Storage Service}
\label{sec:amazons3}

Amazon's Simple Storage Service (S3) \cite{amazons3} is object storage with a simple web interface and RESTful API \cite{richardson2013restful} designed to store and retrieve any amount of data from anywhere on the Internet.  Its design allows it to scale past trillions of objects, each of which can be up to 5 terabytes (TB) in size~\cite{s3faq}.  The usage model of a RESTful API is different from that of a traditional POSIX file system and usually requires the explicit usage of a supporting library or code to allow storage, retrieval, or manipulation of data within the object store; otherwise, it follows the familiar model in which metadata are stored and retrieved separately from data.  

Buckets and objects are the primary resources managed within Amazon's S3 and can be likened to folders and files within a traditional general-purpose file system.  Each of these has associated "subresources", such as access control lists (ACLs), versioning information, logging configuration, and life-cycle management information, all of which are handled in much the same way as metadata, such as extended attributes would be on a POSIX file system~\cite{s3usingobjects}.

The Amazon S3 security model is based on policies defined by access control lists attached to the aforementioned buckets or objects as subresources~\cite{amazons3acl}.

\section{Experimental Environment}
\label{sec:environment}

\subsection{Hardware and Operating System}
\label{sec:hardware}

All three of the experiments described below were performed on the MIT SuperCloud~\cite{reuther2013llsupercloud} and the Lincoln Laboratory TX-Green Supercomputer. 

\begin{figure}[ht]
	\centering
	\includegraphics[width=8.75cm,keepaspectratio]{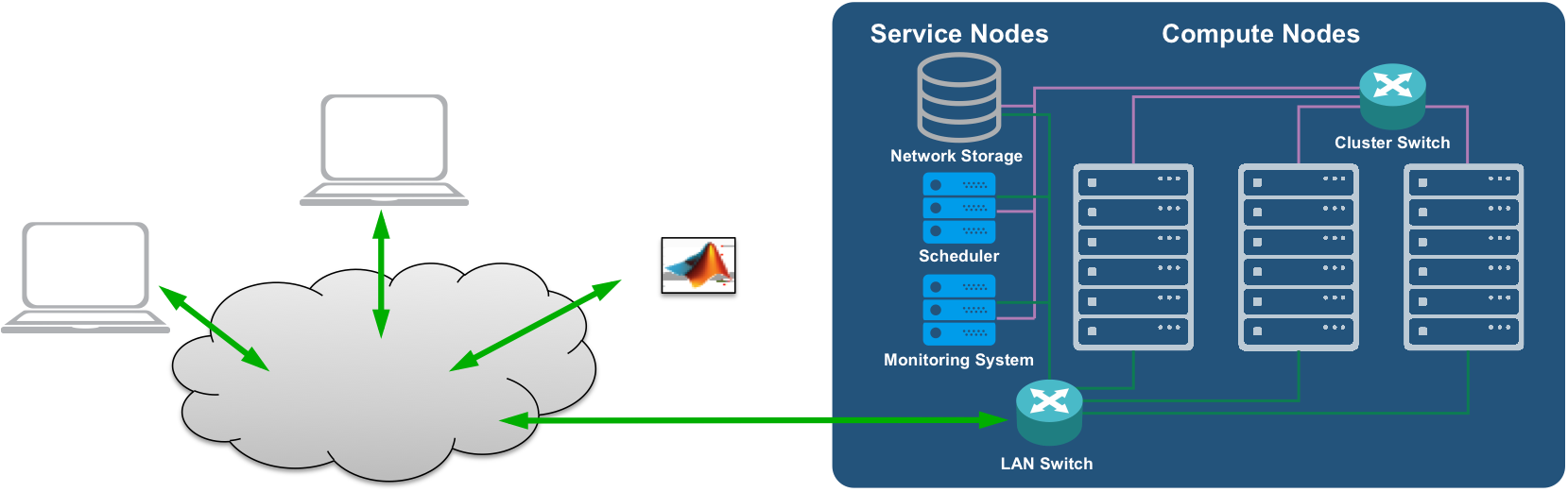}
	\caption{Architecture of the MIT SuperCloud systems.  Users connect to the system over either a local area network or a wide area network.  At the time of connection, their system joins the MIT SuperCloud and can act as a compute node in order to run parallel programs interactively.  The centerpiece of the MIT SuperCloud is several file systems (Seagate, DDN, Dell, Hadoop, and Amazon S3) running on several different network fabrics (10 GigE, InfiniBand, OmniPath).  The MIT SuperCloud is a well-suited platform for performing comparative filesystem benchmarks.}
	\label{fig:supercloudarch}
\end{figure}

The TX-Green Supercomputer at Lincoln Laboratory used for the Lustre Ethernet test is a petascale system that consists of a heterogeneous mix of AMD, Intel, and Knights Landing-based servers connected to a single, non-blocking 10 Gigabit Ethernet Arista DCS-7508 core switch.  All of the compute nodes used in both the single-client and multiple client runs were Haswell-based Intel Xeon E5 servers with 256 GB of RAM.  The Lustre central storage system uses a 5 petabyte Seagate ClusterStor CS9000 storage array that is directly connected to the core switch, as is each individual cluster node.  This architecture provides high bandwidth to all the nodes and the central storage, and is depicted in Figure~\ref{fig:supercloudarch}.

The Lustre Infiniband test made use of the MIT SuperCloud system, a similarly heterogeneous environment sitting atop a 1 petabyte Seagate ClusterStor central storage array connected to a Mellanox FDR network.  Both of these environments run the same version of the LLGrid Operating System Image, a custom Linux distribution based on Fedora Core 20.

To test the performance of data ingest into and retrieval from Amazon Web Services, a single node from the MIT SuperCloud cluster described above was split off, and a SolarFlare SFC9020 10 Gigabit Ethernet card was installed with a direct 10 Gigabit Ethernet fiber link into the Massachusetts Green High Performance Computing Center (MGHPCC) core network.  This regional network provides 0.5 terabits of bandwidth in a loop to participating members, including a 100 gigabit run into a carrier hotel in New York City, which also serves to connect an Amazon Web Services \cite{amazonaws} point of presence. 

\subsection{Software and Test Methods}
\label{sec:software}

As both of the systems under test are network-based storage, albeit of differing types, our experiment calculated the average rate of data transfer based on the number of bytes transmitted and received on the Ethernet and Infiniband interfaces of the client nodes during the period that the transfer occurred.  For the series of single compute node tests, we varied the number of client "worker" processes assigned to copying data while keeping one worker assigned to the storage or retrieval of one 10 GB chunk of data in each case.  In the larger, multi-node, whole-cluster benchmark, a 30 GB blob of data was assigned to each node.

All of the chunks of data stored and received were generated using the pseudorandom number generator provided by the \textit{/dev/urandom} device in Linux in order to avoid the possibility that any underlying compression or deduplication features enabled in the Linux kernel VFS layer, the Lustre file system, or Amazon's Simple Storage Service could skew the results of our test.  In addition to this, to avoid client-side caching of the data obtained in our Lustre read test, a full flush of the Linux kernel page cache (Listing~\ref{lst:dropcaches}) was performed on each compute node prior to each test run in any of the Lustre read tests.  As the Amazon S3 test was performed entirely using the Amazon Web Services CLI tools (\textit{awscli}) using their RESTful API directly and not by means of a file system VFS layer (e.g., FUSE~\cite{miklos2005fuse}), no such steps were required.

\begin{code}[label={lst:dropcaches},caption={Dropping Linux Kernel page cache to avoid VFS layer caching of Lustre read tests.}]
echo "Dropping filesystem page cache.."
echo 1 > /proc/sys/vm/drop_caches
\end{code}

For the Infiniband benchmark to a Lustre file system, the 64-bit counters in the \textit{PortXmitData} and \textit{PortRcvData} fields were queried using the \textit{perfquery} (Listing~\ref{lst:perfquery}) command, with the result multiplied by the number of Infiniband lanes present in the port in order to obtain an accurate value for total bytes transferred. In our test, the link used was a Mellanox Fourteen Data Rate (FDR) Infiniband~\cite{mellanox2015fdr} connection consisting of four lanes of 14 Gb/s each for a maximum aggregate throughput of 56 Gb/s.

\begin{code}[label={lst:perfquery},caption={Using perfquery to get 64-bit Infiniband interface counters.}]
perfquery -x
\end{code}

To capture the Ethernet traffic statistics for both the Lustre on Ethernet and Amazon S3 benchmarks, the output of the "Bytes Transmitted" (1) or "Bytes Received" (9) field in \textit{/proc/net/dev} associated to the appropriate network interface from which the data would be sent or received was sampled once, when all of the worker processes for the test had begun, and then again once the last worker process had terminated.  Both the Amazon S3 and Lustre file system tests were done with a 10 Gigabit Ethernet connection.

\begin{code}[label={lst:batchscript},caption={Bash script to launch a batch of 10 worker processes on a single compute node, each retrieving data from Lustre.}]
for run in {1..10}
do
    echo "Launching process for ${run}..."
    cp ${LOCATION}/randomdata-10G-${run} \
       /dev/null &
    sleep 0.1
done
wait
\end{code}

By measuring traffic statistics at the interface level throughout our test, we were able to monitor for and eliminate the effect of any file system read caching skewing that portion of our experimental results.

Finally, a simple bash script was used to bring it all together, the central loop of which is depicted in Listing~\ref{lst:batchscript}.  Exotic data migration tools were not used for any of the tests; the standard UNIX \textit{cp} file-copy tool was by itself sufficient to achieve adequate performance.  Multiple runs were performed in each test case, with time and interface statistics being sampled and reset at the beginning and end of each run.  The resulting data were examined and aberrant results, likely caused by the variability of the Internet in the Amazon S3 test or concurrent loads causing contention on the shared file system in the Lustre case, were discarded.

\section{Performance Results}
\label{sec:performance}

The benchmark was designed to maximize data ingest and retrieval speed with minimal ramp-up time.  Standard tools were used in a manner generally representative of a typical user workload for copying data.  Processes in the single-client node tests were launched with 100ms of delay between each startup, and, in the large-scale test, the entire run was initiated concurrently.  

\begin{figure}[ht]
	\centering
	\includegraphics[width=8.75cm,keepaspectratio]{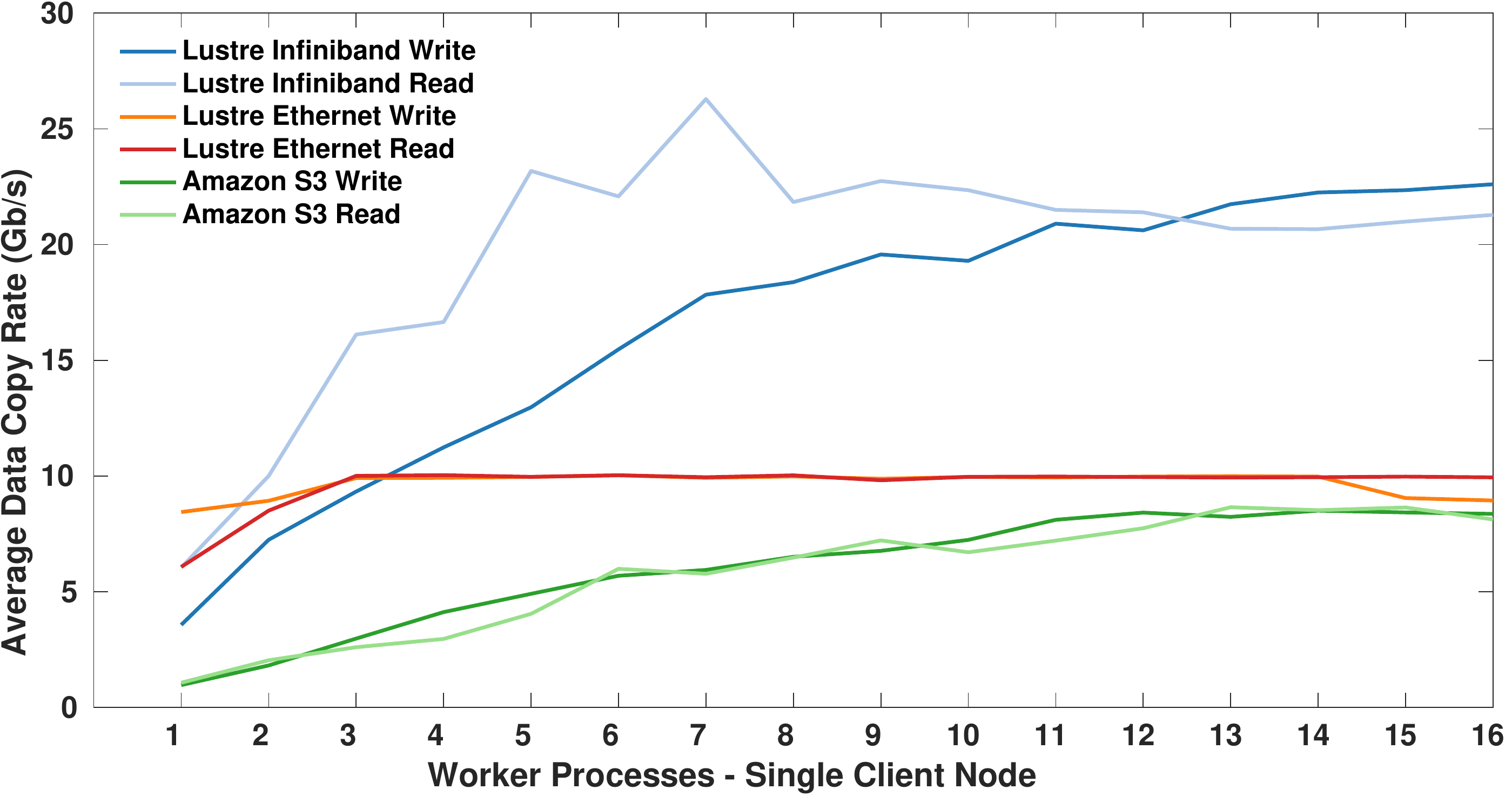}
	\caption{Lustre file system on Infiniband/10 Gigabit Ethernet, and Amazon S3 on 10 Gigabit Ethernet performance as a function of data transfer rate achieved per number of worker processes dispatched.}
	\label{fig:singleperformance}
\end{figure}

Results for the single-client-node run are depicted in Figure~\ref{fig:singleperformance}.  

On our 10 Gigabit Ethernet-based Lustre system, wire-speed read and write performance was achieved with as few as 2 \textit{cp} worker processes, and a single process was able to achieve 70-80\% of peak.  Lustre over Infiniband exhibited a similar performance curve, albeit with a higher potential line rate that is due to the increased potential speed of the Infiniband interface.  

While data transfer to and from Amazon S3 was able to reach the same performance levels as the Lustre file system on our single 10 Gigabit Ethernet test node, it required 12 or 13 separate worker processes to realize peak performance.  A single instance of the AWS copy command (\textit{aws s3 cp}), with no modifications to the Python source, is able to achieve a transfer rate in either direction of approximately 130 MB/s on our test system before consuming 100\% of the single CPU it's running on.  

Using system-level profiling on the running Python process handling the data transfer revealed that data were being read from and written to both the local disk and the network in 8 KB chunks.  This buffer size is much too small to achieve peak performance on anything beyond a very low-bandwidth network.  Further examination reveals that these block and socket buffer size values appear to be hard-coded within the system Python libraries themselves; an example is the the \textit{HTTPConnection.send} method in the system Python \textit{httplib} library, which has a socket buffer size of 8192 explicitly defined as a constant~\cite{pythongithub}.  Given the extreme level of CPU-boundedness displayed by these tools when running at high network rates, it's possible that the Python interpreter itself is also contributing to the high levels of resource utilization, as much of the HTTP protocol and network code in \textit{httplib} is written in pure Python and not bound to a library written in a language such a C or C++.

\begin{figure}[ht]
	\centering
	\includegraphics[width=8.75cm,keepaspectratio]{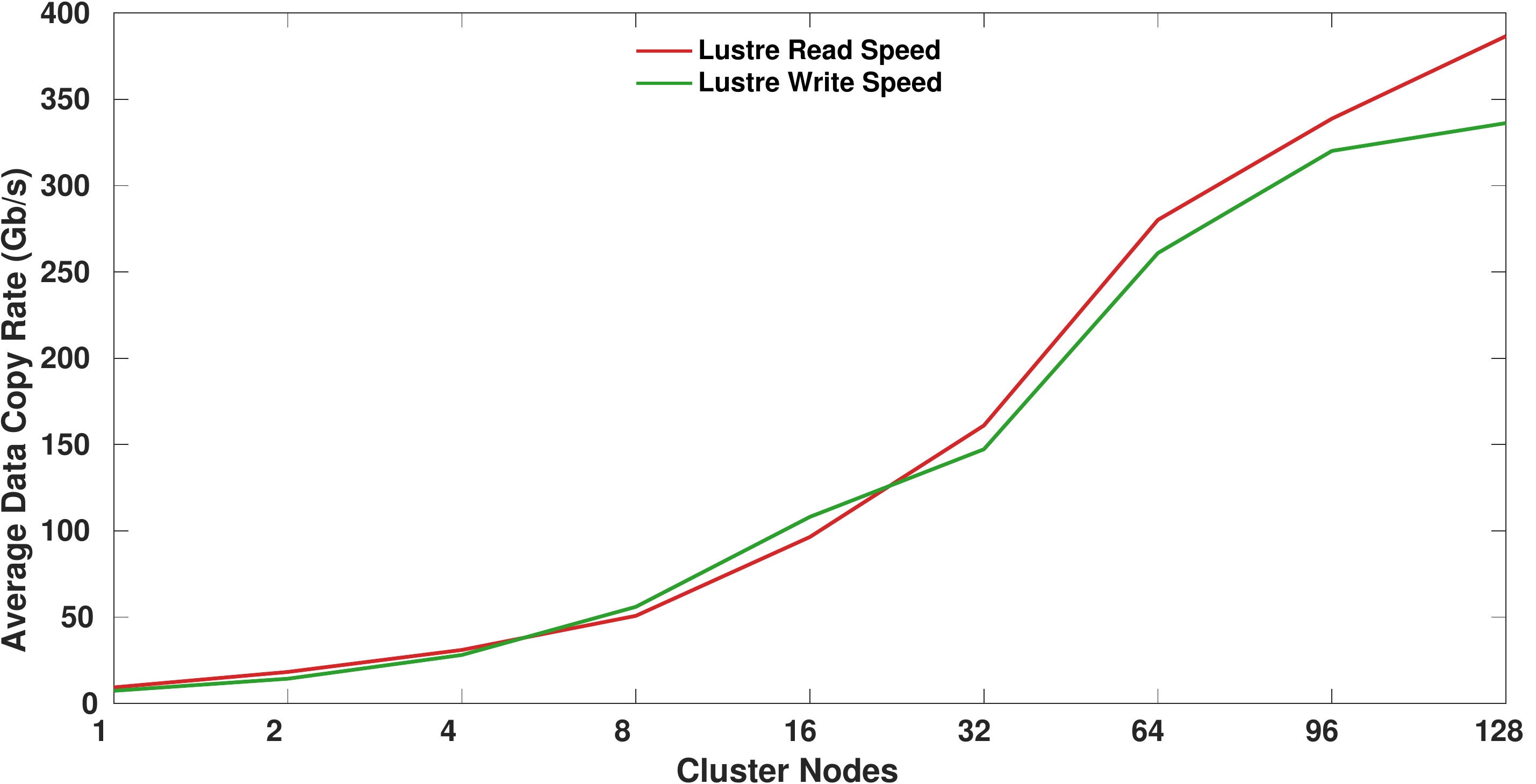}
	\caption{Lustre file system performance scaling on TX-Green Supercomputer using multiple concurrently active 10 Gigabit Ethernet client nodes.}
	\label{fig:multiperformance}
\end{figure}

In the multi-client benchmark results displayed in Figure~\ref{fig:multiperformance}, we demonstrate that the Lustre file system, on modern storage hardware and with a well-designed network architecture, obtains near-linear performance improvement by increasing the number of connected clients retrieving data from or pushing data to the central storage array until the underlying physical limitations of the network and storage hardware are met.  While the results shown in the graph above represent the average start-to-finish transfer rate of each complete test, a sustained peak throughput of 480 Gb/s in the read test and 350 Gb/s in the write test was routinely achieved on the SuperCloud hardware during the 128-node cluster run. 

\section{Summary and Future Work}
\label{sec:summary}

The rise of machine learning and graph analytic systems has created a need for diverse high performance storage and ways to measure and compare the capabilities of these storage systems.  The Lustre file system and Amazon's Simple Storage Service are both designed to address the largest and most challenging data storage problems.  Relatively few comparative measurements exist to inform decisions about which storage systems are best suited for particular tasks.  This paper provides a baseline assessment of the performance and capabilities that can be expected when choosing a storage solution.

The performance tests that we used span the gamut of parallel I/O scenarios ranging from single-client, single-node Amazon S3 and Lustre performance to a large-scale multi-client test designed to demonstrate the capabilities of a modern storage appliance under heavy load.   These results show that when parallel I/O is used correctly (i.e., many simultaneous read or write processes), full network bandwidth performance is achievable and ranged from 10 gigabits/s over a 10 GigE S3 connection to 0.35 terabits/s using Lustre on a 1200-port 10 GigE switch.  These results demonstrate that S3 is well-suited to sharing vast quantities of data over the Internet, while Lustre is well-suited to processing large quantities of data locally.

We have established that one can achieve a very similar baseline-level performance when sequentially reading and writing large objects on both traditional supercomputing network storage such as Lustre and Amazon's cloud-based storage solution.  Traditional general-purpose file systems exposing a POSIX API do provide slightly better ease of use for rapid prototyping purposes; however, there are efforts such as s3fs-fuse~\cite{rizuns3fs} that attempt to bridge that usability gap by providing a file system-like interface to Amazon S3. 

Future work in this area will include further scaling of the Amazon S3 testing beyond a single Internet-connected client node and additional whole-cluster Lustre testing as trays are added to our central storage array.

\section*{Acknowledgments}
\label{sec:acknowledgments}

The authors wish to acknowledge the following individuals for their contributions: Trung Tran, Tom Salter, Paul Burkhardt, Joe Flasher, Chris Hill, Michael Hurley, Anna Klein, Edward Kao, Matt McCarthy, Lauren Milechin, Sanjeev Mohindra, David Martinez, Julie Mullen, Steve Pritchard, Jed Sundwall, Michael Wolfe, and Chuck Yee.

\bibliographystyle{unsrt}
\IEEEtriggeratref{7}
\bibliography{IEEEabrv,references.bib}

\end{document}